\documentclass[apsrev,twocolumn,showpacs,preprintnumbers,amsmath,amssymb]{revtex4}

\usepackage{graphicx}
\usepackage{dcolumn}
\usepackage{bm}

\begin{document}


\title{Infrasound generation by turbulent convection}

\author{M. Akhalkatsi}
\affiliation{Physics Department, Tbilisi State University, Tbilisi
0179, Georgia}

\author{G. Gogoberidze}%
\email{gogober@geo.net.ge} \affiliation{Center for Plasma
Astrophysics, Abastumani Astrophysical Observatory, Tbilisi 0160,
Georgia}

\author{P. J. Morrison}
\email{morrison@physics.utexas.edu} \affiliation {Department of
Physics and Institute for Fusion Studies, The University of Texas
at Austin, Austin, TX 78712}

\date{\today}

\begin{abstract}
Low frequency acoustic wave generation is studied taking into
account the effect of stratification, inhomogeneity of background
velocity profile and temperature fluctuations. It is shown that
for the typical parameters of convective storms the dipole
radiation related to temperature inhomogeneities is at least of
the same order as radiation of Lighthill's quadrupole source. It
is also shown that the source related to stratification could have
valuable contribution whereas some other sources are shown to be
inefficient.
\end{abstract}

\maketitle

\section{\label{sec:1}Introduction}               

It is long known that strong convective storms, such as supercell
thunderstorms are powerful sources of infrasound
\cite{CYH60,BB71,GH}. Detailed observations of convective storm
generated infrasound \cite{BB71,NT03} provides that at least two
different group of infrasonic signals could be identified. The
first group, with characteristic period about $1$ s, has been
found to be in strong connection with proto-tornadic structures,
funnel clouds and tornadoes. Based on coincident radar
measurements of tornados, which show strong relationship between
funnel diameter and infrasound frequency, it is usually supposed
that these infrasound waves are generated by radial vibrations of
the funnel core \cite{BG00}. The second group of infrasound
signals has the periods from $2$ to $60$ s. Usually the emission
appears about 1 hour before observation of tornado. These waves
are not related with tornado itself and are caused by convective
processes that precedes tornado formation. The acoustic power
radiated by convective storm system could be as high as $10^7$
watts \cite{G88}. Although several reasonable mechanisms have been
suggested to explain this acoustic radiation, the physical
mechanism of the process remains unexplained \cite{GG75,G88,BG00}.

Broad and smooth spectrum of the observed infrasound radiation
indicates that turbulence is one of the promising sources of the
radiation. Lighthill's acoustic analogy \cite{L52} represents the
basis for understanding of the generation of sound by turbulent
flows. In this approach the flow is assumed to be known and the
sound field is calculated as a small by-product of the flow.
According to this theory in the case of uniform background
thermodynamic parameters interaction of turbulent vortices
provides quadrupole source of sound. The acoustic power of the
source was estimated by Proudman \cite{P52}. But usage of this
estimation for the infrasound radiation from convective storms
usually leads to the underestimation of the acoustic power
\cite{GG75,GH}.

It is also well known that any kind of inhomogeneity of the
background flow, such as stratification, shear of the background
velocity profile and temperature fluctuations leads to appearance
of additional (mainly dipole) sources of sound \cite{S67,H01}.
Resent developments in this direction led to the formulation of
the generalized acoustic analogy \cite{G02} that implies: (i)
dividing the the flow variables into their mean and fluctuating
parts; (ii) substracting out the equation for the mean flow; (iii)
collecting all the linear terms on one side of equations and the
nonlinear terms on the other side; (iv) treating the latter terms
as as known source of sound.

In the presented paper we consider acoustic radiation from
turbulent convection taking into account mentioned above effects
of stratification, inhomogeneity of velocity profile and
temperature fluctuations. Performed analysis shows that for the
typical parameters of supercell storms the dipole radiation
related to temperature inhomogeneities is at least of the same
order as radiation of Lighthill's quadrupole source. It is also
shown that the source related to stratification could have
valuable contribution whereas some other sources are shown to be
very inefficient.

The paper is organized as follows: simple background flow model is
constructed in Sec. \ref{sec:2}. Linear equation governing the
propagation of sound in the flow is obtained in Sec. \ref{sec:3}.
Sources of acoustic radiation are obtained and analyzed in Sec.
\ref{sec:4}. Summary is given is Sec. \ref{sec:5}

\section{\label{sec:2} Background flow model}

In this section we construct simplified model for background
updraft flow. Suppose in the stratified dry atmosphere there
exists some region (see Fig. 1) with supply of relatively hot air
at some background level and a sink at some height $h$. Assume
pressure, density and temperature fields outside this region are
$P_0(z), \rho_0(z)$ and $T_0(z)$ respectively. In the case of
isentropic atmosphere \cite{GH}
\begin{equation}
P_0(z)=P_{0b}\left( 1-\frac{\gamma -1}{\gamma} \frac{z}{H}
\right)^{\gamma/(\gamma-1)}, \label{eq:21}
\end{equation}
where $H\equiv R T_{0b}/g$ is stratification length scale;
$P_{0b}$ and $T_{0b}$ are pressure and temperature at the
background level; $z$ is vertical coordinate; {\bf g} is gravity
acceleration and $\gamma \equiv c_p/c_v$ is adiabatic index.

In the case of isothermal atmosphere instead of Eq. (\ref{eq:21})
we have
\begin{equation}
P_0(z)=P_{0b}\exp\left( -z/H \right), \label{eq:22}
\end{equation}

Suppose the temperature of the supplied hot air is
$T_{1b}>T_{0b}$. The continuity, Euler, heat and state equations
governing the stationary adiabatic updraft motion in the region
are:
\begin{equation}
{\bf \nabla} \rho_1 {\bf V_1}=0,\label{eq:23}
\end{equation}
\begin{equation}
\rho_1 \left( {\bf \nabla} {\bf V_1} \right) {\bf V_1}+ {\bf
\nabla}P_1-\rho_1 {\bf g}=0,\label{eq:24}
\end{equation}
\begin{equation}
P_1d\rho_1-\gamma \rho_1 dP_1=0,\label{eq:25}
\end{equation}
\begin{equation}
P_1=\rho_1 R T_1,\label{eq:26}
\end{equation}
where ${\bf V}_1={\bf V}+{\bf V}_D$; ${\bf V}(z)\equiv [0,0,V(z)]$
is updraft motion velocity; $z$ is vertical coordinate; ${\bf
V}_D({\bf r})=[\alpha(z)x,\alpha(z)y,0]$ is horizontal velocity
associated with divergence of updraft motion; and $\rho_1(z),
P_1(z), T_1(z)$ are density, pressure and temperature
respectively.

The boundary condition at the boundary of regions 1 and 2 implies
\begin{equation}
P_1(z)=P_0(z).\label{eq:27}
\end{equation}
Eq. (\ref{eq:23}) provides for $\alpha(z)$
\begin{equation}
\alpha(z)=\frac{\partial_z(\rho_1 V)}{2\rho_1},\label{eq:28}
\end{equation}

For other variables straightforward calculations yield
\begin{equation}
\rho_1(z)=\rho_{0b} \frac{T_{0b}}{T_{1b}} \left(
1-\frac{\gamma-1}{\gamma} \frac{z}{H}
\right)^{1/(\gamma-1)},\label{eq:29}
\end{equation}
\begin{equation}
T_1(z)={T_{1b}} \left( 1-\frac{\gamma-1}{\gamma} \frac{z}{H}
\right),\label{eq:210}
\end{equation}
\begin{equation}
V^2(z)=2gz\left(\frac{T_{1b}}{T_{0b}}-1 \right) ,\label{eq:211}
\end{equation}
for isentropic atmosphere and
\begin{equation}
\rho_1(z)=\rho_{0b}\frac{T_{0b}}{T_{1b}} \exp\left(
-\frac{z}{\gamma H} \right), \label{eq:212}
\end{equation}
\begin{equation}
T_1(z)={T_{1b}} \exp\left[ -\frac{(\gamma-1)z}{\gamma H}
\right],\label{eq:213}
\end{equation}
\begin{equation}
V^2(z)=2g \left[ \frac{T_{1b}}{T_{0b}} \frac{\gamma H}{\gamma -1}
\left(1-\exp\left( -\frac{\gamma-1}{\gamma} \frac{z}{H} \right)
\right) - z \right] ,\label{eq:214}
\end{equation}
for isothermal atmosphere.

Eqs. (\ref{eq:28}) and (\ref{eq:211}) yields
\begin{equation}
\frac{V}{|V_D|} \sim \frac{H}{L},\label{eq:215}
\end{equation}
where $L$ is horizontal length scale of region 1. Therefore if $H
\gg L$ we conclude that
\begin{equation}
V\gg |V_D|.\label{eq:216}
\end{equation}

Considered model of background flow do not include horizontal
rotation of the updraft flow as well as horizontal wind with
vertical shear. Existence of the latest one is known to be
necessary for the formation of the strongest convective storm
systems, supercell storms \cite{K87,B}. However, as it will be
shown below none of these flows can have valuable direct influence
on the infrasound generation.

\begin{figure}[t]
\includegraphics[width=\hsize]{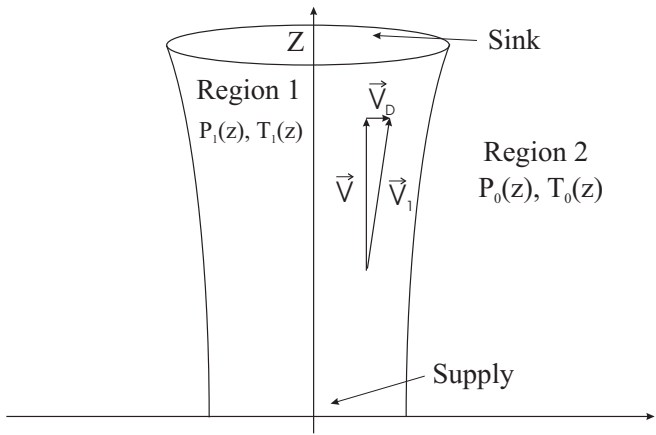}
\caption{Schematic illustration of the considered updraft
flow model.} \label{fig:fig1}
\end{figure}

\section{\label{sec:3}linear operator for acoustic waves}

Next step of the study in the framework of mentioned above
generalized acoustic analogy is derivation of the linear equation
for acoustic wave propagation. Taking into account Eq.
(\ref{eq:216}) linearized continuity, Euler and state equations
can be written as
\begin{equation}
D_t\left( \frac{\rho}{\rho_1}\right) + {\bf \nabla} (\rho_1 {\bf
v})=0,\label{eq:31}
\end{equation}
\begin{equation}
{\bar D}_tv_x +\partial_x\frac{p}{\rho_1}=0,\label{eq:32}
\end{equation}
\begin{equation}
{\bar D}_tv_y +\partial_y\frac{p}{\rho_1}=0,\label{eq:33}
\end{equation}
\begin{equation}
D_t v_z + v_z\partial_z V + \frac{1}{\rho_1} \left(
\partial_z p - \frac{\partial_z P_1}{\rho_1}\rho \right)=0,\label{eq:34}
\end{equation}
\begin{equation}
\frac{p}{P_1} -\gamma \frac{\rho}{\rho_1} =0,\label{eq:35}
\end{equation}
where ${\bf v}, \rho$ and $p$ are perturbations of velocity,
density and pressure respectively; $D_t \equiv
\partial_t + V\partial_z V$ is convective derivative; and ${\bar
D}_t \equiv D_t+\alpha$.

Combining Eqs. (\ref{eq:31})-(\ref{eq:35}) for perturbation of
total enthalpy
\begin{equation}
Q \equiv \frac{p}{\rho_1}+v_z V,\label{eq:36}
\end{equation}
we obtain following equation
\begin{equation}
{\hat L}Q=0,\label{eq:370}
\end{equation}
where ${\hat L}$ is so-called third order linear wave operator
\begin{equation}
{\hat L} \equiv {\bar D}_t D_t \frac{1}{c_s^2}D_t -{\bar D}_t
\frac{\partial_z\rho_1\partial_z}{\rho_1} - D_t
\nabla_\perp^2=0,\label{eq:37}
\end{equation}
and $c_s\equiv (\gamma P_1/\rho_1)^{1/2}$ is sound velocity and
$\nabla_\perp^2\equiv\partial_x^2+\partial_y^2$.

Operators similar to ${\hat L}$ are well known in jet noise theory
\cite{Go76,H01}. Obtained linear equation describes not only the
acoustic wave propagation but it also describes the instability
wave solutions that are usually associated with large scale
turbulent structures and continuous spectrum solutions that are
related to "fine-grained" turbulent motions \cite{G02,Go84}. In
the presence of any kind of inhomogeneity (in the case under
consideration all the background variables depend on vertical
coordinate $z$) linear coupling between these perturbations is
possible, and in principle acoustic waves can be generated by both
instability waves and continuous spectrum perturbations. But in
the case of low Mach number ($M\equiv V/c_s \ll 1$) flows both
kind of perturbations are very inefficient sources of sound. The
acoustic power is proportional to $e^{-1/2M^2}$ and $e^{-\pi
\delta/2 M}$ for instability waves and continuous spectrum
perturbations respectively \cite{CH90,??04}. In the last
expression $\delta$ is the ratio of length scales of energy
containing vortices and background velocity inhomogeneity
($V/\partial_zV$). In the case of supercell thunderstorm $M\sim
0.1-0.15$ and $\delta \sim 10^{-2}$, therefore both linear
mechanisms have negligible acoustic output and attention should be
payed to sources of sound related to nonlinear terms and entropy
fluctuations that will be studied in the next section.

Propagation of infrasound in the atmosphere was intensively
studied by different authors ( see, e.g., \cite{OGCG04,K04} and
references therein) and will not be considered in this paper.

\section{\label{sec:4} sources of infrasound}

For obtaining the sources of acoustic waves generated by turbulent
convection we follow standard technique of acoustic analogy.
Dividing the flow variables into their mean and fluctuating parts
keeping the nonlinear terms in Eqs. (\ref{eq:31})-(\ref{eq:34}),
and noting that in general in the flow there can exist entropy
fluctuations, so replacing Eq. (\ref{eq:35}) by
\begin{equation}
\frac{p}{P_1} -\gamma \frac{\rho}{\rho_1} =
\frac{s}{c_v},\label{eq:41}
\end{equation}
where $s$ is entropy fluctuation, after straightforward
calculations we obtain following equation
\begin{equation}
{\hat L}Q=S_l+S_s+S_{V}+S_T,\label{eq:42}
\end{equation}
where
\begin{equation}
S_l=\partial_t[\partial_x\partial_i(v_iv_x)+
\partial_y\partial_i(v_iv_y)]+ {\bar D}_t \left[
\partial_z - \frac{V}{c_s^2}D_t \right]\partial_i(v_iv_z) ,\label{eq:43}
\end{equation}
\begin{equation}
S_s={\bar D}_t
\partial_i(v_iv_z) \left[ \frac{\partial_z\rho}{\rho} - V^2
 \partial_z \left( \frac{1}{c_s^2}\right) \right],\label{eq:44}
\end{equation}
\begin{equation}
S_V=-{\bar D}_t \frac{V\partial_zV}{c_s^2}
\partial_i(v_iv_z),\label{eq:45}
\end{equation}
\begin{equation}
S_T= \frac{{\bar D}_t}{c_p} \left[ \partial_tD_ts +
\frac{\partial_z(s\partial_zP_1)}{\rho_1} -
\frac{V\partial_z\rho_1}{\rho_1}\partial_ts-
\frac{\partial_z(sV^2\partial_z\rho_1)}{\rho_1}
\right],\label{eq:46}
\end{equation}
are source terms. In the derivation of Eq. (\ref{eq:42}) all the
nonlinear terms containing $\rho v_i$ has been omitted, due to the
fact that these terms are known to be much weaker sources of sound
then the terms containing $v_i v_j$ in the low Mach number flows
\cite{Go76}.

All the sources presented in Eq. (\ref{eq:42}) are well known in
aeroacoustics. Consequently, we do not repeat calculations of
acoustic power  (i.e., acoustic energy generated in a unit time)
of each source term and adopt well known results
\cite{Go76,H01,S67,GMK94} for the problem under consideration. The
main idea of mentioned methods of acoustic power calculations is
the following: in the low Mach number flows convective propagation
effects have minor influence on sound generation process and
therefore all the terms containing $V$ on the left hand side of
Eq. (\ref{eq:42}) can be neglected. This circumstance allows one
to obtain simple estimations of acoustic power of the sources.

In the absence of background flow ($V=0$) the first term $S_l$
reduces to the Lighthill's quadrupole source \cite{Go76}. Its
acoustic power
\begin{equation}
N_l \sim \frac{\rho \bar v^8}{lc_s^5}F,\label{eq:47}
\end{equation}
where $\bar v$ is rms of turbulent velocity fluctuations and $F$
is total volume of turbulent flow. This expression was used in
ref. \cite{GG75} to estimate acoustic power of supercell storm.

The term $S_s$ is dipole source caused by the stratification. The
influence of stratification on acoustic wave generation has been
studied intensively in the context of solar physics
\cite{S67,GMK94}. Related acoustic power
\begin{equation}
N_s \sim \frac{\rho {\bar v}^6 l}{c_s^3 H^2}F=
\frac{c_s^2}{\omega^2H^2}N_l, \label{eq:48}
\end{equation}
where $\omega={\bar v}/l$ is the characteristic frequency of
emitted acoustic waves. Substituting $c_s=340$ m/s, $H=10^4$ m and
$\omega \sim 0.1~{\rm s^{-1}}$, we obtain $N_s \sim 0.1 N_l$,
i.e., the power of this source is weaker then acoustic power of
Lighthill's quadrupole source, but in principle considered dipole
source could have valuable contribution to the total acoustic
power.

The source term $S_V$ is related to inhomogeneity of updraft
velocity. It is dipole source similar to $S_s$. Corresponding
acoustic power
\begin{equation}
N_V \sim \frac{\rho l V^4 \bar v^6}{H^2 c_s^7}F=M^4N_s,
\label{eq:49}
\end{equation}
is negligibly small in comparison with $N_s$ for low Mach number
flows (for supercell storms $M\sim 0.1-0.15$).

The last term $S_T$ is caused by entropy fluctuations. The most
strong acoustic source of this kind is so-called thermo acoustical
source \cite{H01} that is related to small scale inhomogeneities
of background density (and therefore temperature) fluctuations,
that produce dipole source. The physics of this kind of acoustic
radiation is the following: "hot spots" or "entropy
inhomogeneities" behave as scattering centers at which dynamic
pressure fluctuations are converted directly into sound. The
acoustic power
\begin{equation}
N_T \sim \frac{\rho \Delta T^2 \bar v^6}{l T^2 c_s^3}F,
\label{eq:410}
\end{equation}
where $\Delta T$ denotes the rms of temperature fluctuations.

Taking for supercell storm \cite{B} $\bar v \sim (3-10)$ m/s;
$\Delta T \sim (3-10)^\circ$K and $T=270^\circ$ K we see that for
the typical parameters of supercell storm the dipole radiation
related to temperature inhomogeneities is at least of the same
order as radiation of Lighthill's quadrupole source.

In the considered model no attention was payed to the horizontal
wind with vertical shear and rotation of the updraft flow. These
assumptions allow us to obtain relatively simple equations. These
flows could not make valuable changes in the acoustic output
estimations. Indeed, consideration of the wind and rotation can
lead to appearance in the right hand side of Eq. (\ref{eq:42})
additional dipole terms similar to $S_V$ that has negligible
acoustic radiation, and also additional quadrupole source term
known as "shear noise" \cite{Go76}. Acoustic power of the later
one is also much weaker compared to Lighthill's quadrupole source
in low Mach number flows.

In the presented paper we study dry atmosphere and therefore
influence of humidity on the sound generation in the convective
storms has not been considered. We intend to study this influence
in the future.

\section{\label{sec:5}summary}

The simple model of the background flow has been constructed and
the problem of acoustic radiation from turbulent convection taking
into account the effects of stratification, inhomogeneity of
velocity profile and temperature fluctuations has been considered
in the framework of acoustic analogy. Performed analysis shows
that for the typical parameters of supercell storms the dipole
radiation related to temperature inhomogeneities is at least of
the same order as radiation of Lighthill's quadrupole source. It
is also shown that the source related to stratification could have
valuable contribution whereas dipole and quadrupole sources
related to the inhomogeneity of background velocity profile are
shown to be very inefficient.

\begin{acknowledgments}

Authors are grateful to Ming Xue for valuable help.

The research described in this publication was made possible in
part by Award No. 3315 of  the Georgian Research and Development
Foundation (GRDF) and the U.S. Civilian Research and Development
Foundation for the Independent States of the Former Soviet Union
(CRDF). The research was supported in part by ISTC grant No. G
553.

\end{acknowledgments}

\end{document}